\documentclass[twocolumn]{aastex63}
\usepackage{multirow}
\usepackage{amsmath,amsfonts}

\newcommand\msun{$M$\mbox{$_{\normalsize\odot}$}}

\newcommand\lbol{$L$\mbox{$_{\rm bol}$}}

\newcommand\Teff{$T_{\rm eff}$}

\newcommand\logl{$\log (L/L_\odot)$}

\newcommand\halpha{H${\sc \alpha}$}

\newcommand\N[1]{{\color{blue} \bf #1}} %N8

\usepackage{soul}

\received{June 1, 2019}
\revised{January 10, 2019}
\accepted{\today}
\submitjournal{ApJ}
\shorttitle{The Yellow Supergiants of Wd1}
\shortauthors{Beasor, Smith \& Andrews }
\graphicspath{{./}{figures/}}
\begin{document}

\title{Don't believe the hype(r): The Yellow Supergiants of Westerlund 1}

\email{embeasor@gmail.com}

\author{Emma R. Beasor}\altaffiliation{Hubble Fellow}
\affiliation{Steward Observatory, University of Arizona,
933 N. Cherry Ave., Tucson, AZ 85721, USA}
\affiliation{NSF's NOIR Lab,
950 N. Cherry Ave., Tucson, AZ 85721, USA}
\author{Nathan Smith}
\affiliation{Steward Observatory, University of Arizona,
933 N. Cherry Ave., Tucson, AZ 85721, USA}
\author{Jennifer E. Andrews}
\affiliation{Gemini Observatory, 670 N. Aohoku Place, Hilo, Hawaii, 96720, USA}

%\nocollaboration{5}

%% Note that the \and command from previous versions of AASTeX is now
%% depreciated in this version as it is no longer necessary. AASTeX 
%% automatically takes care of all commas and "and"s between authors names.

%% AASTeX 6.3 has the new \collaboration and \nocollaboration commands to
%% provide the collaboration status of a group of authors. These commands 
%% can be used either before or after the list of corresponding authors. The
%% argument for \collaboration is the collaboration identifier. Authors are
%% encouraged to surround collaboration identifiers with ()s. The 
%% \nocollaboration command takes no argument and exists to indicate that
%% the nearby authors are not part of surrounding collaborations.

%% Mark off the abstract in the ``abstract'' environment. 
\begin{abstract}
Yellow hypergiants (YHGs) are often presumed to represent a transitional post-red supergiant (RSG) phase for stars $\sim$30-40 \msun. Here we present visual-wavelength echelle spectra of six YHG candidates in the Galactic cluster Westerlund 1, and we compare them to known YHGs, IRC +10420 and Hen3-1979. We find that the six YHG candidates do not exhibit any metallic emission lines, nor do they show strong H$\alpha$ emission, and as such do not meet the criteria necessary to be classified as YHGs. In conjunction with their moderate luminosities of \logl = 4.7-5.4 estimated from optical/infrared photometry, we suggest instead that they are normal yellow supergiants (YSGs) with more modest initial masses around 15-20 \msun. This adds additional support to the hypothesis that Wd1 is a multi-age cluster with an older age than previously assumed, and is not a $\sim$5 Myr old cluster caught at a very specific transitional point when single-star evolution might yield Wolf-Rayet stars, luminous blue variables (LBVs), RSGs, and YHGs in the same cluster. Nevertheless, the population of YSGs in Wd1 is very unusual, with YSGs outnumbering RSGs, but with both spanning a large luminosity range. Here, we discuss evolutionary scenarios that might have led to the high fraction of YSGs. The number of YSGs and their significant luminosity spread cannot be explained by simple population synthesis models with single or binary stars. Even with multiple ages or a large age spread, the high YSG/RSG ratio remains problematic. We suggest instead that the objects may experience a prolonged YSG phase due to evolution in triple systems. 
\end{abstract}

%% Keywords should appear after the \end{abstract} command. 
%% See the online documentation for the full list of available subject
%% keywords and the rules for their use.

%% From the front matter, we move on to the body of the paper.
%% Sections are demarcated by \section and \subsection, respectively.
%% Observe the use of the LaTeX \label
%% command after the \subsection to give a symbolic KEY to the
%% subsection for cross-referencing in a \ref command.
%% You can use LaTeX's \ref and \label commands to keep track of
%% cross-references to sections, equations, tables, and figures.
%% That way, if you change the order of any elements, LaTeX will
%% automatically renumber them.
%%
%% We recommend that authors also use the natbib \citep
%% and \citet commands to identify citations.  The citations are
%% tied to the reference list via symbolic KEYs. The KEY corresponds
%% to the KEY in the \bibitem in the reference list below. 

\section{Introduction}
Yellow hypergiants (YHGs) are an extremely rare class of object, which, in a simplified single-star evolutionary paradigm, are presumed to represent a brief post-red supergiant (RSG) phase for some of the most massive stars with initial masses of 30-40 \msun\ \citep[e.g.][]{de1998yhg}. Observationally they occupy the `Yellow Void' \citep[$\sim$ 4000 - 1000K, ][]{de1998yhg,gordon2019postrsg} on the Hertzsprung-Russel diagram (HRD) and have high luminosities at or around the Humphreys-Davidson (HD) limit \citep[\lbol\ $\sim$ 5.5, ][]{davies2018humphreys,mcdonald2022red}. Typical yellow supergiants (YSGs) are thought to primarily be stars evolving away from the main sequence (MS) and toward the RSG phase.  In contrast, YHGs have been suggested to arise following a period of extreme mass-loss during the RSG phase, forcing the  star to return to the hotter regions of the HRD\footnote{Though some studies do suggest normal, lower mass YSGs may also experience multiple blue-loops, see e.g. \citep{gordon2019postrsg}}. It is traditionally thought that strong mass-loss in a preceding RSG phase has raised the L/M ratio, leading the star to a precarious instability near the Eddington limit, thus enhancing atmospheric turbulence and mass loss that may explain the strong emission lines observed in these stars \citep{de1998yhg}. A YHG may continue to bounce between the hot and the cool regions, undergoing multiple `blue-loops' \citep[e.g.][]{stothers1968evolution}. During this transition phase, any potential strong mass loss may change the fate of the star and ultimately determine where on the HRD it ends its life, and what kind of supernova (and remnant) will be produced. As such, they potentially represent a significant turning point in the lives of the most massive stars. However, observational estimates of RSG mass-loss rates show that normal RSG winds are too weak \citep{beasor2020mass} to drive stars to become YHGs as envisioned unless one invokes a hypothetical extreme mass-loss phase for RSGs.  Moreover, this interpretation of YHGs as the products of single-star mass loss neglects the observed fact that binaries and multiple systems are very common at high initial masses \citep{moe2017mind}, but in any case, the strong mass loss is likely to significantly influence the end fate of the star.

The confirmation of an object as a true YHG relies on observational factors that merit the Ia+ luminosity class. The most common criteria used are the ``Keenan-Smolinski" criteria, as put forward by \citet{de1998yhg}. In addition to having a high luminosity (\lbol$\geq$ 5.5), a bona-fide YHG will also exhibit one or more broad components of H${\sc \alpha}$ ($\sim$ 100 km/s) and have absorption lines that are significantly broader than those of Ia stars of similar spectral types and luminosities. It is also expected that a YHG will have several low excitation metal emission lines, such as Fe {\sc ii} and Ca {\sc ii} \citep{clark2005massive}. These spectral features are thought to be indicative of the strong mass-loss that had led the YHG away from the RSG phase, as noted above.   There are only a handful of objects which meet these criteria, for example IRC + 10420 \citep{humphreys1979studies}, $\rho$ Cas \citep{lobel2003rho}, Hen3-1979 \citep{lebertre1989friedegg} and HR517A \citep{ches2014hr}. In addition, if the star has been driven to the YHG phase by strong RSG mass loss in the recent past, one might expect to see evidence of this past mass ejection in the form of dense circumstellar shells, strong mid-IR excess from circumstellar dust, or molecular emission. Given the rarity of these objects, and their potentially huge impact on the fates of massive stars, it is important we understand how common they may be throughout the Universe. 

Westerlund 1 (Wd1) is a young massive cluster in the Milky Way that contains a large population of evolved massive stars, including six YHG candidates \citep{clark2005massive, clark2020vlt}. Under the assumptions that Wd1 is a simple stellar population (i.e. all single stars), the stellar diversity in the cluster --- in particular the simultaneous presence of Wolf-Rayet stars (W-Rs), a luminous blue variable (LBV), YHGs, and red supergiants (RSGs) --- can only be explained by a narrow range of ages around 5 Myr \citep[see][]{clark2005massive}. This would require the stars to have very high initial masses ($\geq$ 35\msun) because of the particular post-MS evolutionary tracks of such stars in models that transition to the blue with strong adopted RSG mass loss. In addition to the clues it offers about the relationship between these various evolutionary phases of massive stars, the age of Wd1 would therefore have important implications for the progenitor of the magnetar located in the cluster \citep{muno2007magnetar}.

\citeauthor{clark2005massive} suggested that the luminosities of the YHGs could exceed \logl=5.7 using a scaling relation of the O{\sc i} absorption line. This would place these stars at or above the upper luminosity limit for cool supergiants. Recent works however have shown that the evolutionary history of Wd1 is likely far more complicated \citep[][]{aghakhankoo2019inferring,beasor2021wd1,navarete2022distance,negueruela2022westerlund}. For one, re-appraising the luminosities of the RSGs and YHG candidates showed that the luminosities of the objects are too low to be consistent with a 5 Myr cluster, and instead imply an age of 10 Myr and a significantly less massive progenitor population. However, the presence of a high mass eclipsing binary system (W13) is in tension with the lower luminosity population. Indeed, the stellar diversity of Wd1 cannot be explained by a single-age population of either single star or binary star models, and it seems more likely that Wd1 was formed over a prolonged period of star formation \citep{beasor2021wd1}. 

The downward revision in luminosity for the YHG candidates (with a resulting luminosity range of \logl\ = 4.7 -- 5.22) would suggest that the stars no longer fit the specification of being bona-fide YHGs, as defined in \citep{de1998yhg}, and are more likely normal YSGs transitioning to the red. As such, it is important to determine if their spectra also fail to meet the classification criteria to be YHGs. 

Here, we aim to independently constrain the lower luminosities and hypergiant vs. supergiant classification implied by photometry by investigating spectral indicators of luminosity class using MIKE echelle spectra. In Section \ref{sec:spectra} we present the spectra and compare them to well known YHGs IRC+10420 and Hen3-1979. We find that, indeed, these stars in Wd1 do not have spectra characteristic of YHGs, and instead have spectra consistent with normal moderate-luminosity YSGs. Despite this, they continue to pose a number of problems for stellar evolutionary theory, including the high YSG/RSG number ratio, and the large spread in luminosities. In Section \ref{sec:results} we discuss the results and put them into the context of stellar evolution, and we speculate about the evolutionary scenarios that may have given rise to this unusual stellar population.

\begin{figure*}
    \centering
    \includegraphics[width=18cm]{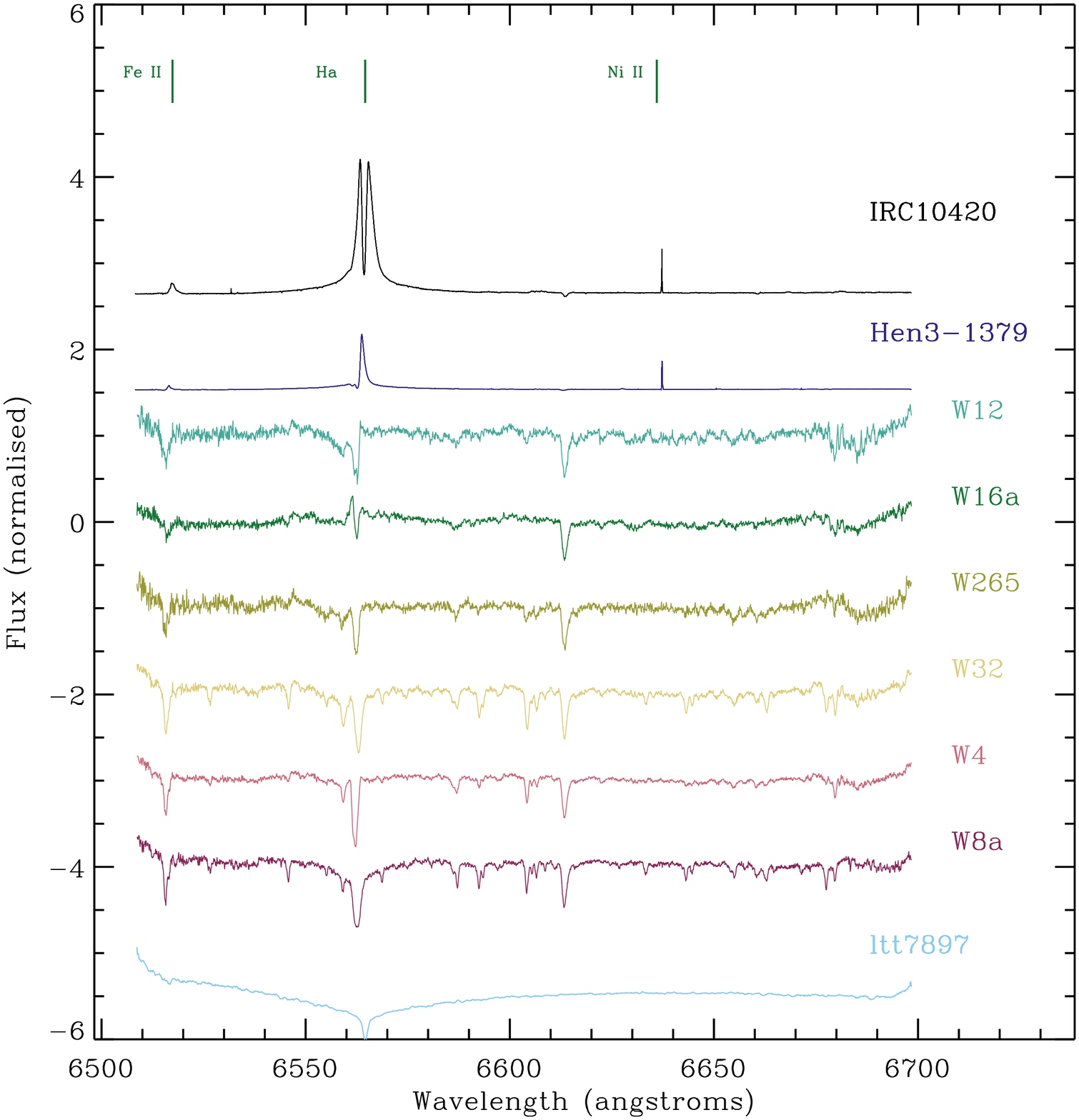}
    \caption{MIKE spectra of the 6 YHG candidates; W12, W16a, W265, W32, W4 and W8a, and the standard star LTT 7897, between 6500 and 6700 ${\rm \AA}$. For comparison, we also show ESO-FEROS spectra of YHGs IRC+10420 and Hen3-1379, and note the positions of prominent emission lines. }
    \label{fig:ord16}
\end{figure*}

\begin{figure*}
    \centering
    \includegraphics[width=18cm]{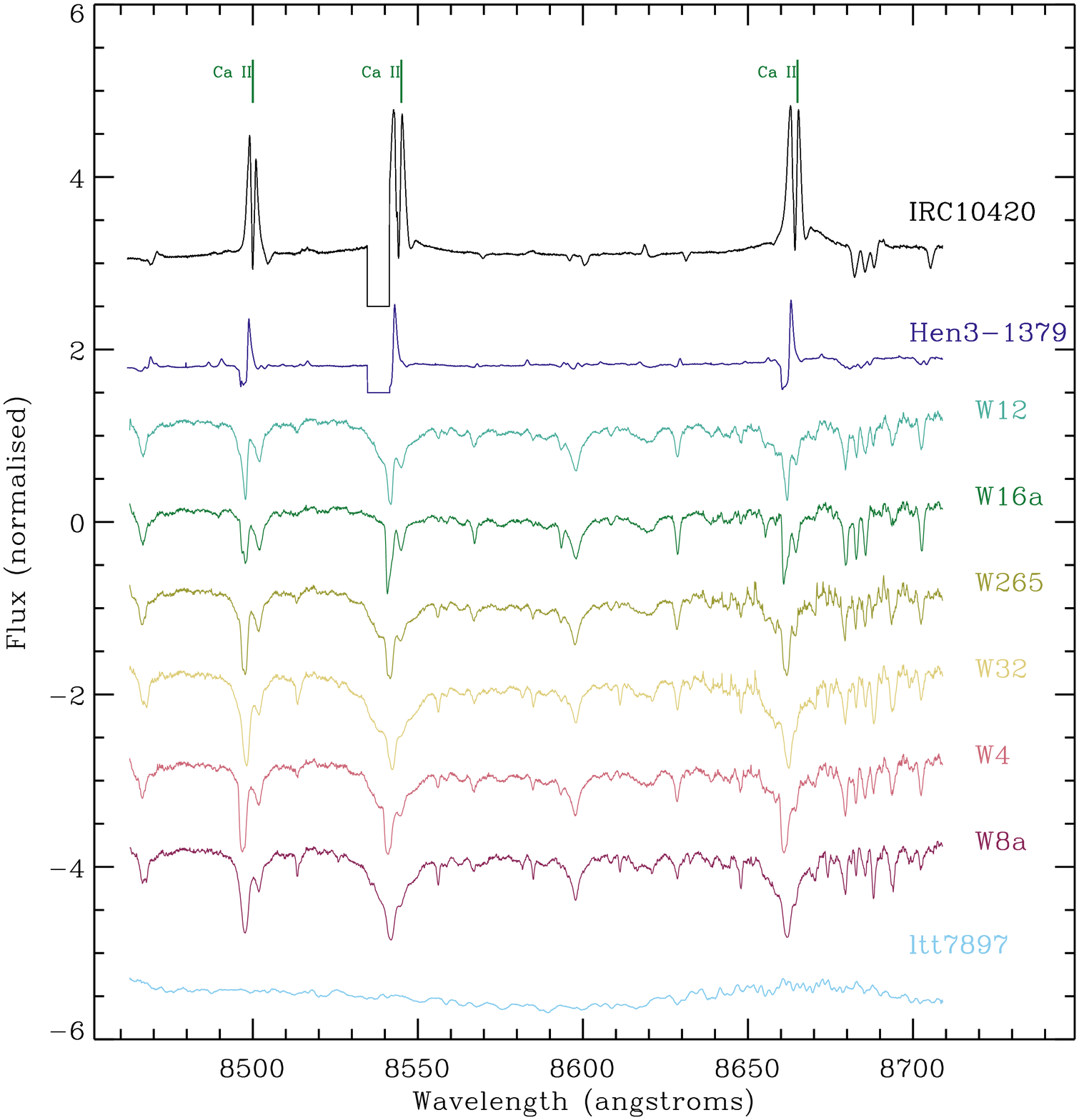}
    \caption{Same as Fig. 1 but for 8400 to 8700 ${\rm \AA}$.}
    \label{fig:ord4}
\end{figure*}

\section{Observations and Data Reduction}\label{sec:spectra}
We obtained optical echelle spectra of the six YHG candidates in Wd1 on 2021 July 31, using the Magellan Inamori Kyocera Echelle (MIKE) spectrograph, which is a double
Echelle spectrograph designed for use at the Magellan Telescopes
at Las Campanas Observatory in Chile \citep{bernstein03}. The observations consisted of 3 $\times$ 300s exposures (except for W16a, for which the exposures were 3 $\times$ 600s) with the 1 arcsec $\times$ 5 arcsec slit. The spectra were reduced using the MIKE pipeline\footnote{https://code.obs.carnegiescience.edu/mike/}
(written by D. Kelson). Although MIKE records the entire optical spectrum in separate blue and red channels, the high interstellar reddening toward Wd1 resulted in a very low signal-to-noise ratio in the blue channel, so we only discuss red portions of the spectra below.

Fig. \ref{fig:ord16} shows the $\lambda\lambda$6550 - 6700 {\AA} region of the spectra for each YHG candidate, as well as archival ESO-FEROS \citep{kaufer1999feros} spectra for IRC +10420 and Hen3-1379 (the Fried Egg Nebula). This wavelength range highlights the location of H$\alpha$ emission. Similarly, Figure~\ref{fig:ord4} shows the region of the spectra around the Ca~{\sc ii} infrared triplet.

\section{Results}\label{sec:results}
\subsection{Emission lines}
\begin{figure*}
    \centering
    \includegraphics[width=\columnwidth]{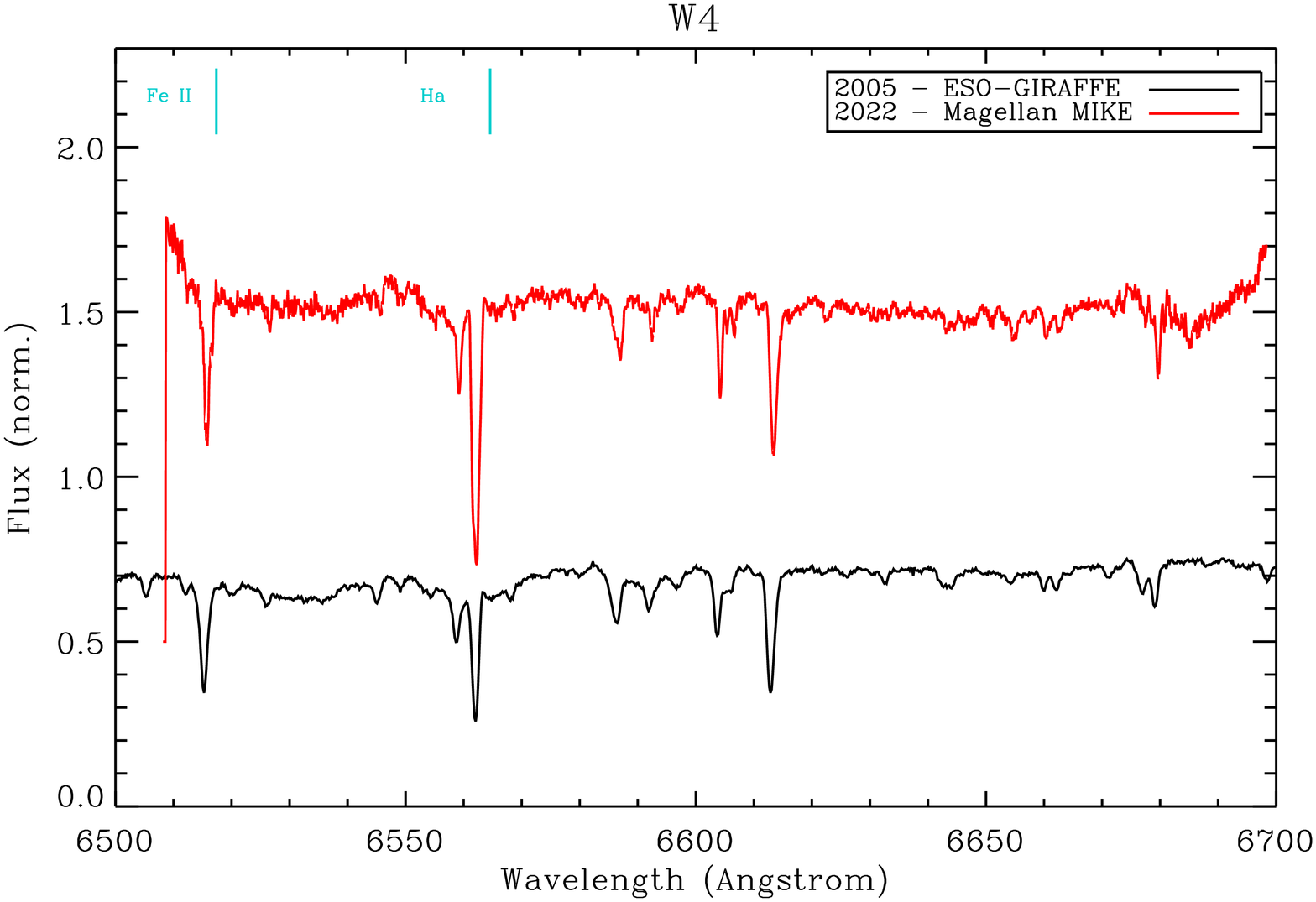}\includegraphics[width=\columnwidth]{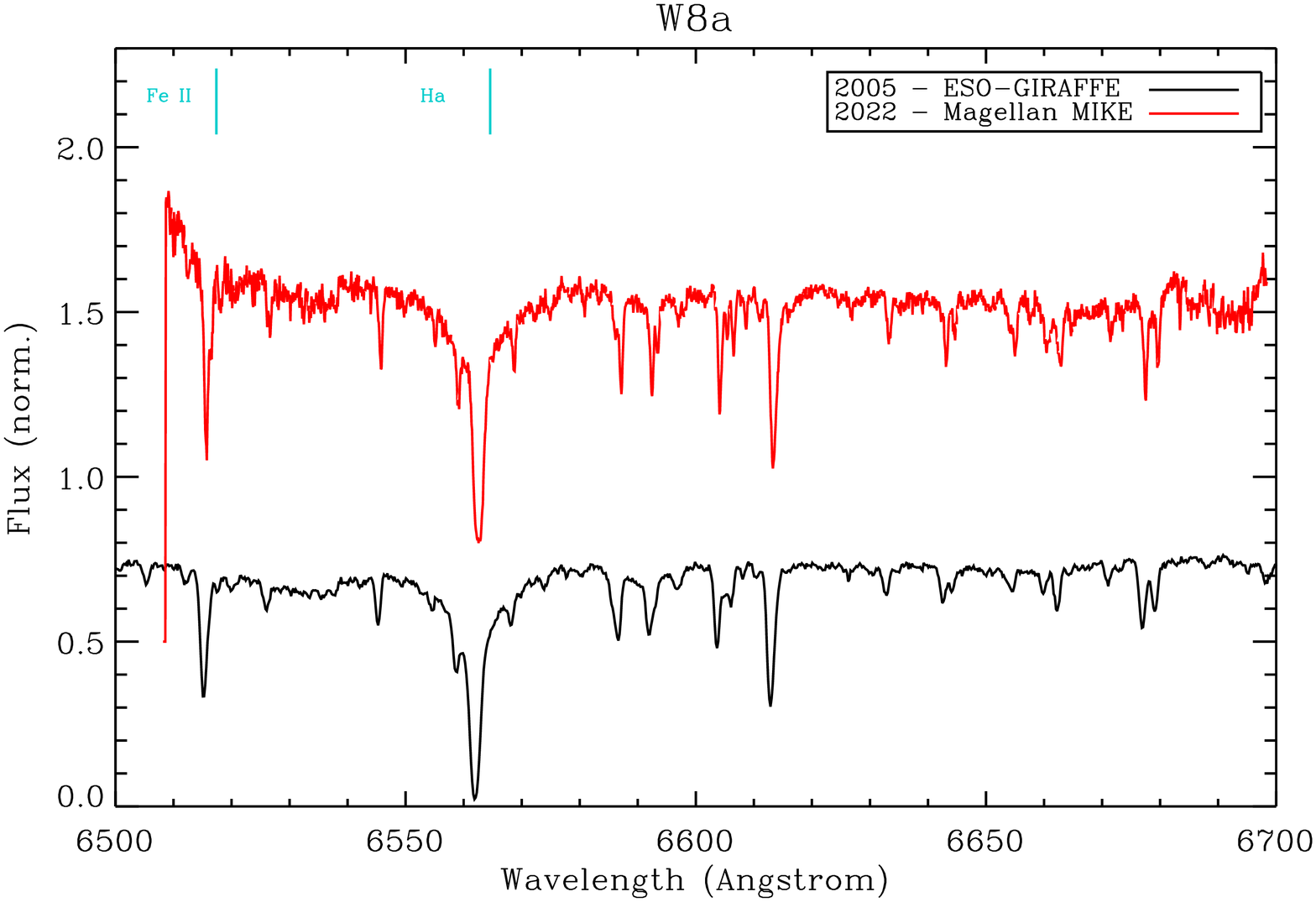}
    \includegraphics[width=\columnwidth]{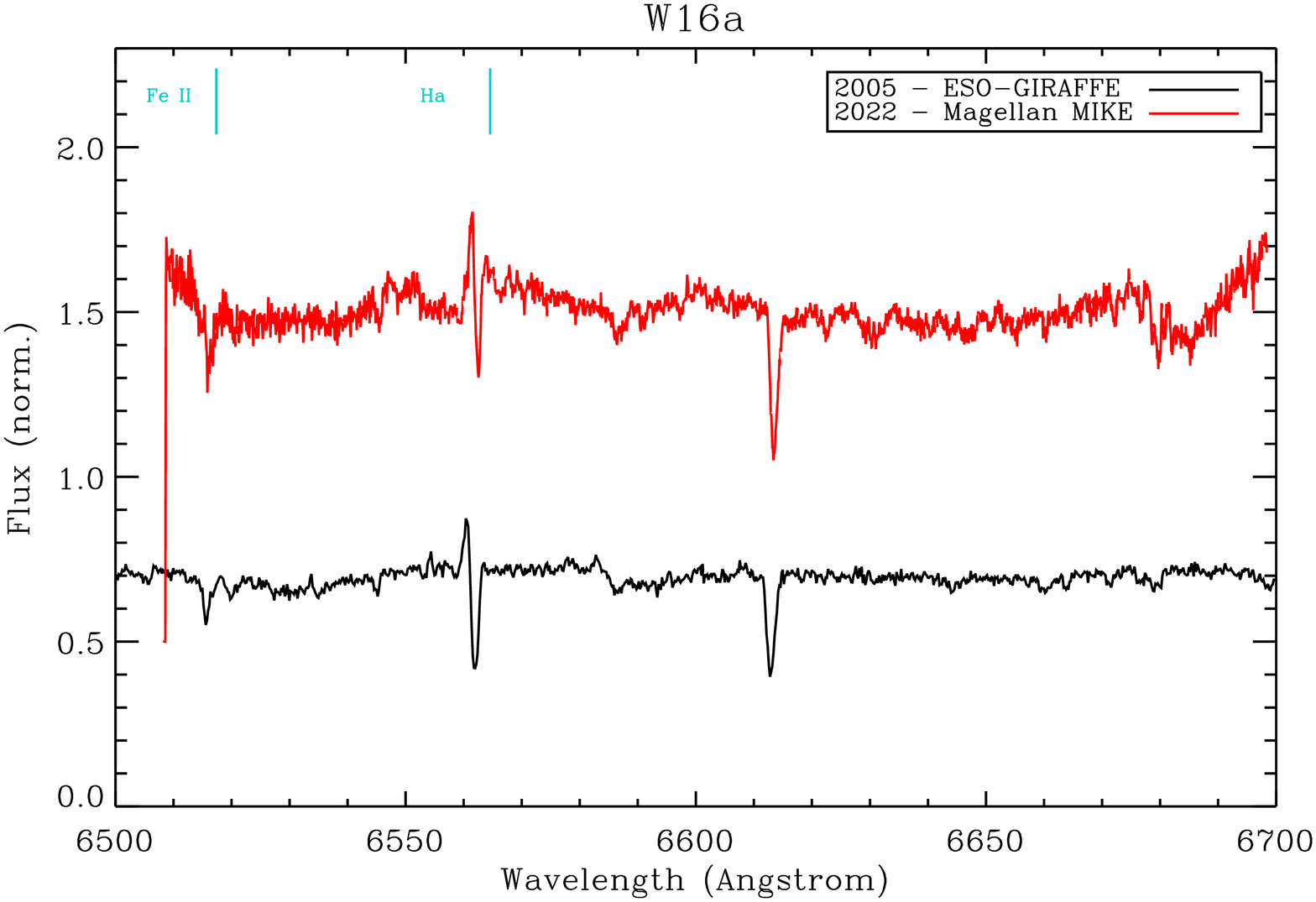}\includegraphics[width=\columnwidth]{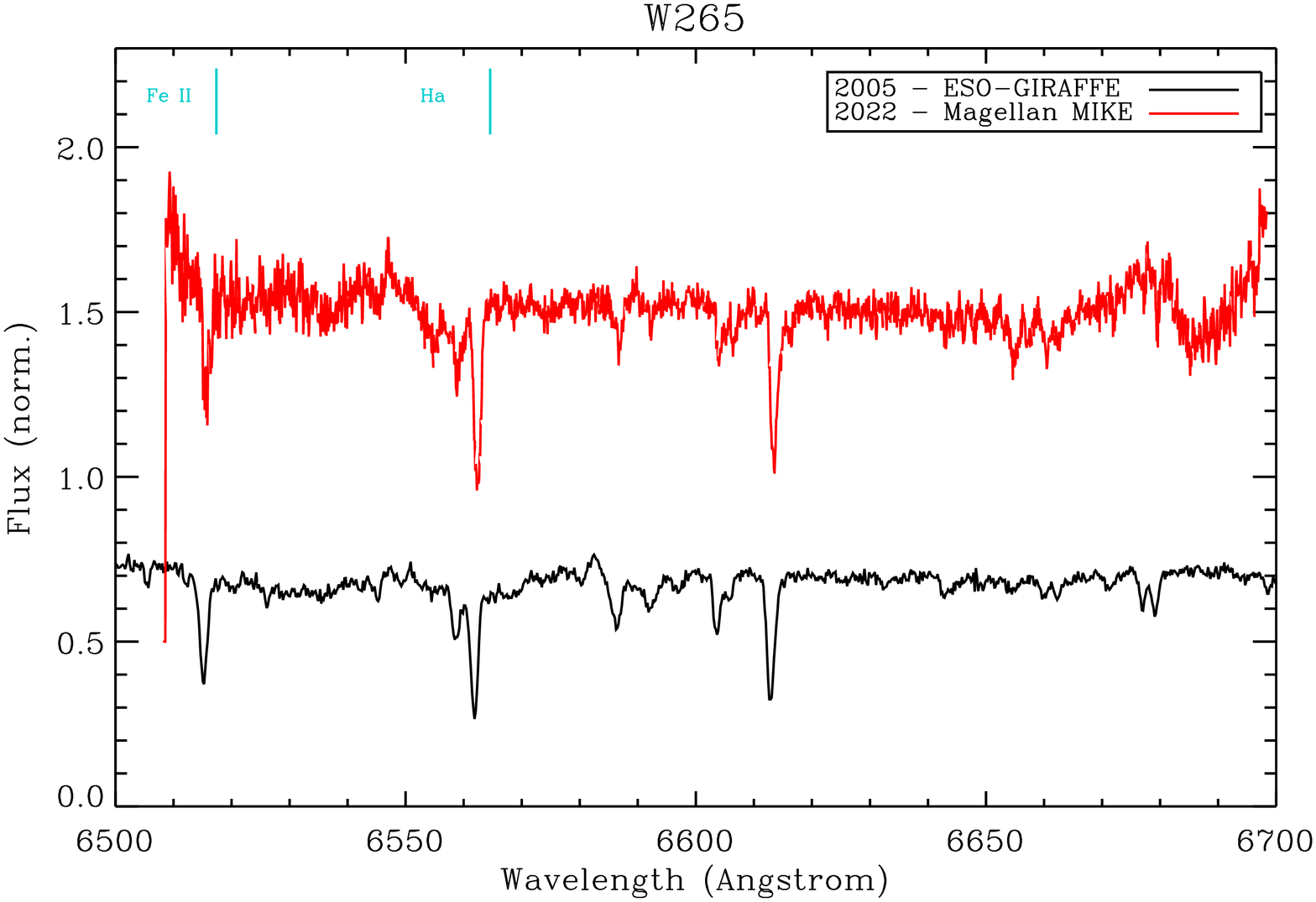}
    \caption{Spectra of W4, W8a, W16a and W265 from 2005 compared to the new data presented in this work. We highlight the position of the H{$\alpha$} line. }
    \label{fig:comparison}
\end{figure*}

Under the Keenan-Smolinski criteria (see above), YHGs should have strong \halpha\ emission and low excitation metal emission lines present, which form in the extreme winds of the stars. As can be seen in Figures \ref{fig:ord16} \& \ref{fig:ord4}, the YSGs in Wd1 do not exhibit the strong H{\sc $\alpha$} and Fe {\sc ii} emission lines seen in IRC +10420 and Hen3-1979.  The only potential exception to this is W16a, which appears to show a slight blueshifted H$\sc{ \alpha}$ emission component at 6560 {\AA}, perhaps indicative of wind emission, although this emission is much weaker than in the two example YHGs. It is also worth noting that W16a is the only YHG candidate in Wd1 to be classified as spectral type A, rather than F, implying it is hotter \citep{beasor2021wd1}. W12 may also have some weak H$\alpha$ emission that partly fills-in the deep absorption. For comparison in Figure \ref{fig:comparison} we also show the 2005 ESO-GIRAFFE spectra presented in \citep{clark2005massive} for 4 of the YSGs for which data was available in the ESO archive. As noted by \citet{clark2010serendipitous} there is a lack of strong emission features in the 2005 spectra. While \citeauthor{clark2010serendipitous} postulated that the objects may have been in a phase of low activity, it seems unlikely that we would observe {\it all} of the objects in a low activity phase at two different epochs. The lack of emission in combination with the reduced luminosities presented in \citet{beasor2021wd1}, demonstrate that these six stars in Wd1 are {\it not} bona-fide YHGs. Instead, they are more likely to be lower-mass YSGs. 

A similar trend is seen in the Ca~{\sc ii} infrared triplet.  Both IRC +10420 and Hen 3-1379 show strong emission, albeit with prominent central self-absorption or P Cygni absorption, whereas the target stars in Wd1 show no net emission in Ca~{\sc ii}.

\subsection{Luminosties from O~{\sc i} }
Previous luminosity estimates for the YSGs in Wd1 were derived from the correlation between the strength of the O{\sc i} line and $M_{\rm V}$ \citep{ferro2003revised}. \cite{clark2005massive} measured the equivalent widths (EWs) of the line for each of the YHGs and assumed a bolometric correction of BC$_{\rm V}$=0 to derive luminosities of \logl\ $\leq$ 5.7, far higher than those found when integrating under the spectral energy distribution \citep[SED,][]{beasor2021wd1}. As discussed in \cite{beasor2021wd1}, this discrepancy could be due to applying the $M_{\rm V}$ - O{\sc i} relation to EWs beyond its calibrated range \citep{arellano2003ow}.  

\begin{figure}
    \centering
    \includegraphics[width=\columnwidth]{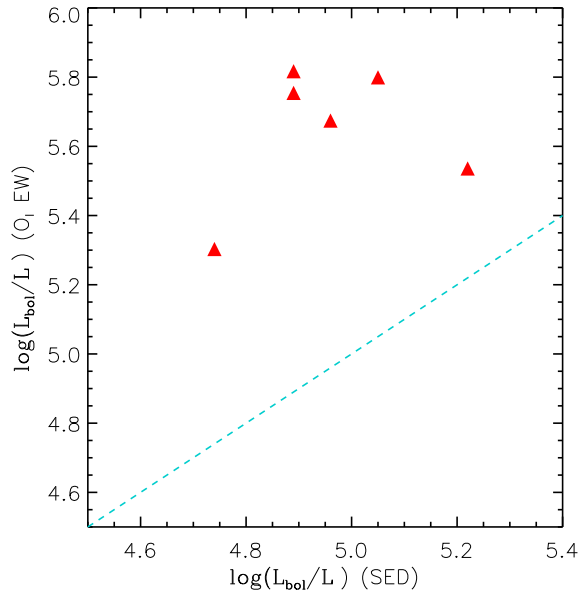}
    \caption{Luminosities derived from the SED and O{\sc i} EW relation for the 6 YSGs in Wd1. A 1:1 relation is shown by the blue dashed line. }
    \label{fig:lbol_ysg_compare}
\end{figure}

With our bolometric luminosities and echelle spectra, we can assess whether or not the $M_{\rm V}$ - O{\sc i} relation holds for these stars. In Table \ref{tab:Oews} we show the EWs for the  O{\sc i} 7774 and Fe{\sc i} 7748 lines derived from the MIKE spectra, as well as the O{\sc i} 7774 EWs measured by \citet{clark2005massive}. Here we use the O{\sc i} relation with the EWs found from the MIKE spectra to calculate $M_{\rm V}$ for each YSG and compare the derived luminosity from the EWs to that found from integrating under the SED \citep{beasor2021wd1}. Figure \ref{fig:lbol_ysg_compare} demonstrates that when using the EWs for the YSGs and the predicted relationship, the derived luminosity is significantly overestimated for all the YSG stars in Wd1. 

\begin{table}[]
    \centering
    \begin{tabular}{|lccc|}
    \hline
    & \multicolumn{2}{c}{This work} & Clark et al. 2005 \\
 
       Star  & O{\sc i} 7774 & Fe{\sc i}  7748 & O{\sc i} 7774 \\
       \hline
        W4 & 2.90$\pm$0.04& 0.05$\pm$0.01& 3.0 \\ 
        W8 & 2.11$\pm$0.06 & 0.22$\pm$0.05&2.1 \\
        W12 & 2.94$\pm$0.04 & 0.015$\pm$ &2.94\\
        W16 & 2.62$\pm$0.05 & - & 2.7 \\
        W32 & 2.40$\pm$0.06 & 0.286$\pm$0.015 & 2.3 \\ 
        W265 & 2.78$\pm$0.05 & 0.07$\pm$0.013 & 2.78 \\
        \hline
    \end{tabular}
    \caption{Equivalent widths for O{\sc i} 7774 \& Fe{\sc i} in each of the YSGs.}
    \label{tab:Oews}
\end{table}

\begin{figure*}
    \centering
    \includegraphics[width=18cm]{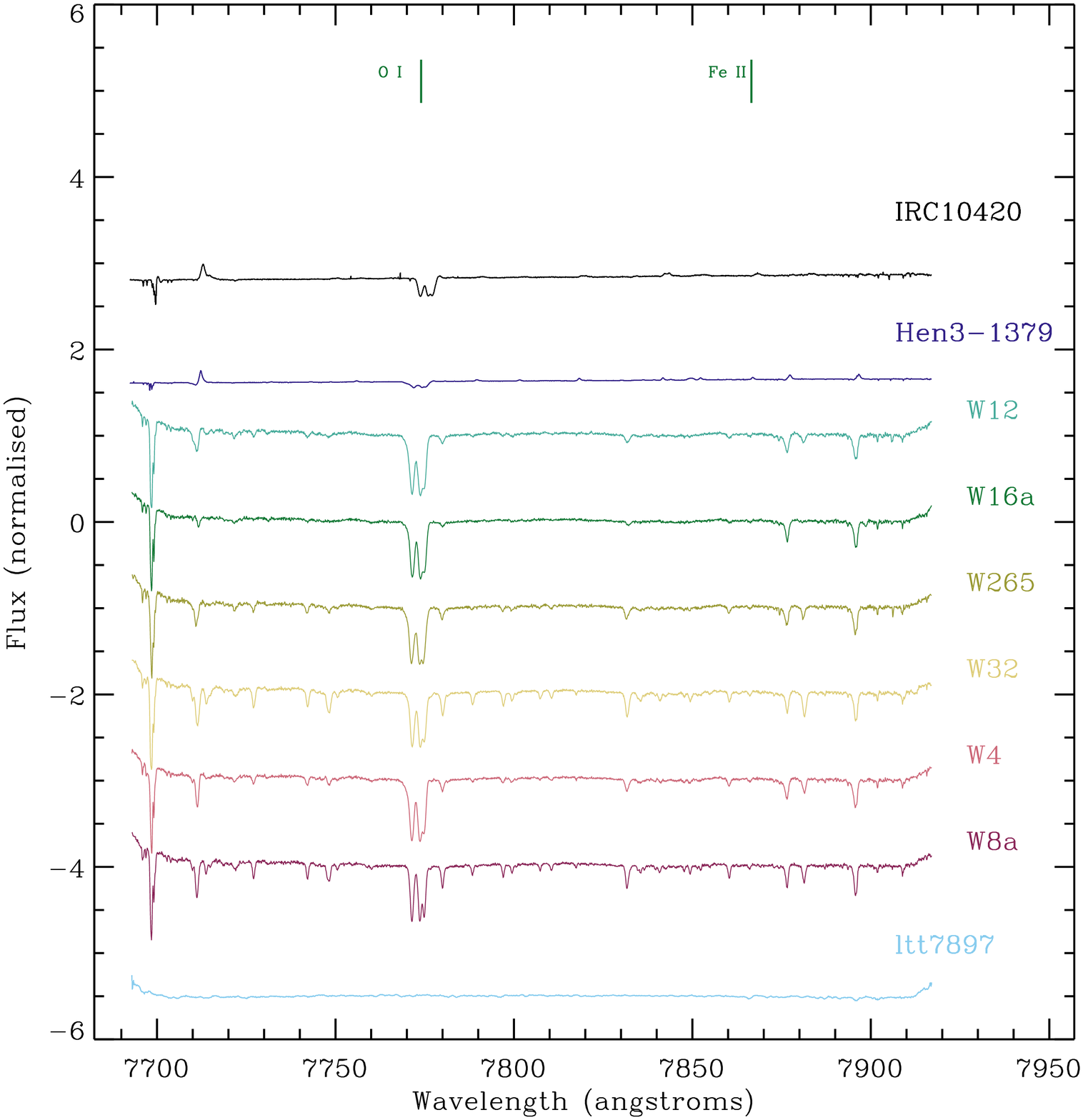}
    \caption{Same as Fig. 1 but for 7700 to 7900 ${\rm \AA}$. The location of the O{\sc i} 7774 line is indicated. }
    \label{fig:ord8}
\end{figure*}

\section{Discussion}
\subsection{The YSGs of Wd1}
Wd1 has long been of interest to the massive star community, due to the presence of a large number of exotic evolved star types, including RSGs, Wolf-Rayet stars, an LBV, a magnetar, and the 6 YHG candidates which we have now demonstrated are normal YSGs, not YHGs). Assuming single-star evolution, the only way all of these objects could co-exist would be if the cluster was at an extremely specific and young age, $\sim$ 5 Myr \citep{clark2005massive}. This can occur only at this age because this marks a special turning point where all of these types of stars are found at different times along the post-MS evolutionary path of stars with approximately the same initial mass around 35-40 \msun\ (at least when using the unrealistically high mass-loss rates that were adopted in older single-star evolutionary codes). Above this mass, stars avoid the RSG phase, and below this mass, they do not turn back to the blue.

The luminosities of the cool supergiants (CSGs) could not be directly measured at the time of the Clark et al.\ study  (due to a lack of usable mid-IR photometry), and instead were inferred from the cluster age and the assumption that Wd1 formed in a single starburst. At 5 Myr, this would imply the RSGs and YSGs to be extremely luminous (\logl\ = 5.5) and hence some of the most massive stars known, with initial masses of $\sim$ 40\msun. More recently, however, it has been suggested that the evolutionary history of Wd1 is more complicated than a single starburst cluster \citep{beasor2021wd1}. Indeed, re-appraising the luminosities of the cool supergiants revealed that they are less massive, inconsistent with being YHGs in the cluster. The lack of strong \halpha\ emission lines in the MIKE spectra presented here support the conclusion that the stars are normal YSGs, as opposed to massive YHGs. Their luminosities inferred directly from photometry, in comparison with single-star evolutionary models, imply lower initial masses of 15-20 \msun, not $\sim$40 \msun\ as inferred previously.

Their demotion from YHG to YSG does not make these stars easy to understand, however. The relatively large number of YSGs present in the cluster is still peculiar, both from a theoretical standpoint and when directly comparing to the observed YSG/RSG number ratios in other young massive clusters. For example, RSGC1 ($\sim$ 10 Myr) contains 16 RSGs, and only 1 YSG. 

Under the single star paradigm, the YSG phase is expected to be extremely short-lived ($\leq$10$^{5}$ yrs for a 10 Myr cluster), while the RSG phase is longer ($\sim$ 10$^{6}$ yrs). As such, one would expect to see more RSGs than YSGs at any given moment in a single age cluster. It was suggested by \citet{georgy2012yellow} that strong mass-loss rates (between 3 and 10 times higher than the fiducial mass-loss) during the RSG phase could cause stars to return to the YSG phase, thus explaining observed YSG SN progenitors. However, studies since then have shown that mass-loss rates during the RSG phase are in fact {\it lower} than usually implemented in models \citep{beasor2016evolution,beasor2018evolution,beasor2020mass}, and so it is unlikely that an RSG would be able to return to the hotter regions of the HR diagram due to RSG wind mass-loss alone. Even if this were possible, one would need an explanation for why these particular stars in Wd1 experienced much higher mass-loss than other stars of similar initial mass in other star clusters. In addition, if the YSGs in Wd1 were post-RSGs, we would expect to see evidence of previous strong mass-loss events, for example through a large infrared excess \citep[as is seen for IRC+102420, e.g.][]{humphreys1997irc}.  However, the spectral energy distributions for all of the YSGs in Wd1 reveal that they do not have substantial mid-IR excess from massive dust shells \citep[see Fig. 2 within][]{beasor2021wd1}. 

The spread in luminosities for the YSGs is also unusual and very difficult to explain if they all have a similar age. Again, if we assume single-star evolution, it is expected that after stars leave the main sequence they cool quickly across the HRD at an almost constant luminosity \citep[e.g.][]{meynet2000stellar,ekstrom2012grids, choi2016mist}. Given that the YSG lifetime is so short, we would expect that any YSG we are currently observing to have the same initial mass to within $\sim$ 1 \msun, and as such they should have similar (or even identical) luminosities.  However, in Wd1 we see the YSGs actually span a range of 0.5 dex in luminosities, from \logl\ = 4.74 up to 5.22. This is difficult to reconcile with single-star evolutionary theory assuming a single age starburst. If, as suggested in \cite{beasor2021wd1}, Wd1 is the product of a sustained period of star formation, then the YSG luminosities imply an age spread of up to 5 Myr (see Fig. 5 within). However, even this scenario cannot explain the relatively high number of YSGs present in the cluster as compared to its number of RSGs. 

\subsection{Possible explanations }
\begin{figure}
    \centering
    \includegraphics[width=\columnwidth]{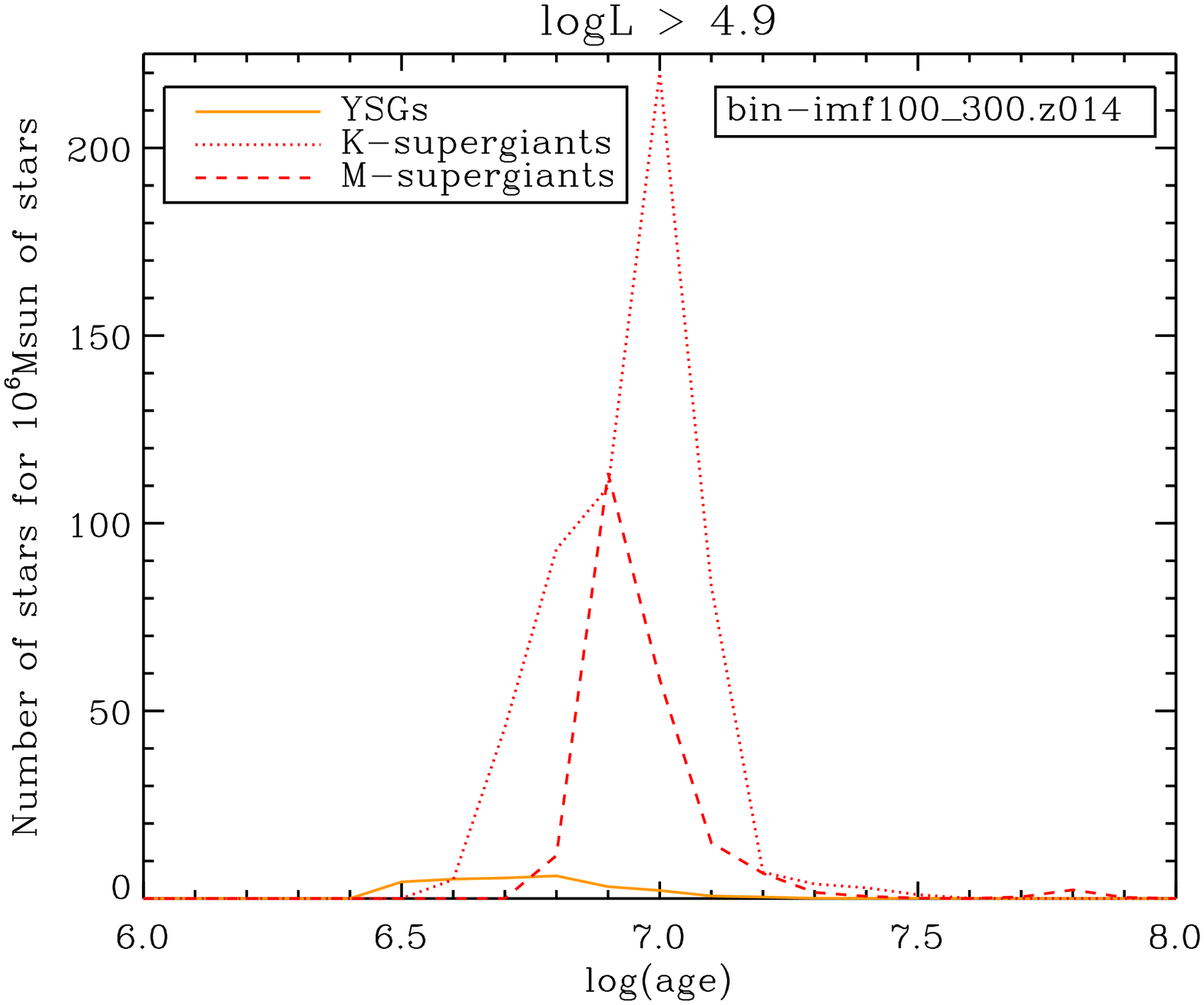}
    \caption{BPASS predicted number of high luminosity stars (\logl\ $>$ 4.9) of given spectral types in a 10$^{6}$\msun\ population. See Table 3 in \cite{eldridge2017bpass} for spectral type definitions.}
    \label{fig:bpass}
\end{figure}

It is clear that the YSG population in Wd1 is difficult to reconcile with our current understanding of evolutionary theory. Both single-star and binary models fail to reproduce fully the range of objects present in the cluster, in particular the high fraction of YSGs and their spread in luminosity. The Binary Population and Spectral Synthesis (BPASS) models provide predicted numbers of stars of a certain spectral type per 10$^6$\msun\ of stars as a function of age (see Fig. \ref{fig:bpass}), for populations including binary and single stars. At Solar metallicity the models predict a YSG:RSG ratio of only 1:20 at $\sim$10 Myr for both sets of models \citep[for the definitions of spectral types see Table 3 in][]{eldridge2017bpass}. Population synthesis with the single star Geneva evolutionary tracks also predicts a lower YSG:RSG ratio than is observed in Wd1. Using the SYnthetic CLusters Isochrones \& Stellar Tracks (SYCLIST) code \citep{georgy2014syclist} we compute a 10 Myr population of single stars with a total mass of 10$^{6}$\msun \footnote{While there is the option to include binaries in SYCLIST, are based entirely of single star models and do not include any form of interaction.}. Using luminosity and temperature cuts to define the RSG vs YSG phase (where an RSG has log(\Teff) $\leq$ 3.6 and \lbol\ $>$ 4.8 and a YSG has log(\Teff) $\geq$ 3.6 and \lbol\ $>$ 4.8, in line with BPASS definitions) we find the YSG-to-RSG ratio is approximately 1:2 in these models.  This is much higher than predicted by BPASS, but still significantly lower than observed in Wd1.

It is now well established that most massive stars exist in binary or even multiple systems \citep[e.g.][]{sana2012binary,moe2017mind}, which can have huge effects on evolution through stripping, or even merging. For RSGs, it has been shown that merging on the MS can lead to the existence of `red straggler stars', which appear more luminous than expected for a group of single stars with a given cluster age \citep[][]{beasor2019discrepancies,britavskiy2019vlt}. This could in part explain the presence of higher luminosity YSGs alongside lower-luminosity neighbors (i.e. different amounts of mass gained through accretion or mergers with a range of companion masses), as they could be evolved merger products on their way to becoming red stragglers. However, this still does not explain the high number of YSGs compared to RSGs. 

Given the lack of mid-IR excess surrounding the YSGs, it seems unlikely that they are post-RSGs. A post-RSG phase is also inconsistent with their relatively large number since in this scenario the YHG phase is a very brief transitional phase as the RSG sheds the last vestiges of its H envelope before evolving to warmer temperatures as a WR star.  Instead, it seems more likely that there is a mechanism that is causing the stars to stall in the warmer YSG phase as they cool and evolve toward the red. \citet{clark2020vlt} noted that the stellar population within Wd1 is heavily influenced by binary interactions, so it is possible that the YSGs themselves have evolved from either binary systems or higher-order multiples. When stars evolve in triple systems the number of potential evolutionary paths increases \citep{toonen2022stellar}, and as such this could provide a potential explanation for the YSG population in Wd1, if indeed Wd1 formed with an unusually high triple fraction. For example, in hierarchical triple systems, the likelihood of a merger event between the two central stars is increased compared to normal binary systems \citep{moe2017mind}. If this occurred, for example on the MS, the merger product would continue to evolve as if it were a single star with a higher initial mass \citep[as is the case for blue and red straggler stars, ][]{beasor2019discrepancies,britavskiy2019red}. If a few such mergers are part of the population in Wd1, it could explain the large spread in YSG and RSG luminosities. However - unique to hierarchical triple systems - the presence of a wide companion that remains after the merger would likely further complicate the onward evolution of that red straggler. At present there is limited work in the literature exploring the effects of triple systems on relative numbers of cool supergiants. For example, if a third star were present in a wide orbit, it may interact with the post-MS merger product as its envelope expands. This stripping of the outermost layers might thwart the envelope's further redward expansion, limiting its photospheric radius to that of a yellow star rather than an RSG. This kind of behaviour may reduce the speed at which a star crosses the yellow void, causing an apparent excess of YSGs. Although speculative, this scenario can be tested by monitoring these stars for radial velocity variations caused by the pull of relatively faint main-sequence companions.  

Another potential solution may arise from the YSGs gaining mass (rather than merging) as they cross the Hertzsprung gap. BPASS models have shown that moderate mass gain on the order of 10 - 20\% may lead to an order of magnitude increase in the lifetime of a YSG, though it is not clear under what conditions this would occur in nature (private communications, J. Eldridge).

\section{Conclusions}
Here we have presented optical echelle spectra of the 6 YHG candidates in Wd1. We show that in combination with their recently revised luminosities, they can be conclusively ruled out as hypergiants.  They are instead normal YSGs of modest luminosity and initial masses in the 15-20 \msun\ range. We also highlight the very high YSG:RSG number ratio in Wd1, and we discuss potential evolutionary scenarios to explain the apparent overabundance of YSGs. We\ find that the high YSG fraction cannot be explained with currently available single star or binary models.

\section*{Acknowledgements}
The authors would like to thank the anonymous referee for useful comments which improved the paper. ERB would like to thank Jan Eldridge and Ben Davies for a number of interesting discussions. ERB is supported by NASA through Hubble Fellowship grant HST-HF2-51428 awarded by the Space Telescope Science Institute, which is operated by the Association of Universities for Research in Astronomy, Inc., for NASA, under contract NAS5-26555. This work makes use of the IDL software and astrolib. 

\bibliography{sample63}
\bibliographystyle{aasjournal}
\end{document}